\documentclass[doublecol]{epl2} 
\usepackage{mathrsfs}
\usepackage{graphicx}
\usepackage{latexsym}
\usepackage{amsmath}
\usepackage{amssymb}
\usepackage{textcomp}
\usepackage{amsbsy}
\usepackage{graphics}
\usepackage{epstopdf}
\usepackage{color}

\title{$f(R)$ gravity with spacetime torsion}

\author{Hitender Kumar \inst{1} \and Tanmoy~Paul\inst{2,3} \and Soumitra~SenGupta\inst{1}}
\shortauthor{H. Kumar, T. Paul, S. Sengupta}

\institute{                    
  \inst{1} School of Physical Sciences, Indian Association for the Cultivation of Science,\\
2A $\&$ 2B Raja S.C. Mullick Road Kolkata - 700 032, India \\
  \inst{2} Department of Physics, Visva-Bharati University, Santiniketan 731235 \\
  \inst{3}Labaratory for Theoretical Cosmology, International Centre of Gravity and Cosmos,
Tomsk State University of Control Systems and Radioelectronics (TUSUR), 634050 Tomsk, Russia\\
}

\abstract{
The duality between a higher curvature $f(R)$ gravity model and a scalar-tensor theory helps to bring out  the role of the additional degree of freedom originating  from the  higher derivative terms in the gravity action. Such a degree of freedom which appears as a scalar field has been shown to have multiple implications in Cosmological/Astrophysical  scenario. The present work proposes a novel generalization to  this correspondence between $f(R)$ gravity and a dual scalar-tensor theory when the affine connection is considered to have an antisymmetric part. It turns out that the $f(R)$ action in presence of spacetime torsion can be recast to a $non-minimally$ coupled scalar-tensor theory with a 2-rank massless antisymmetric tensor field in the Einstein frame, where the scalar field gets coupled with the antisymmetric field through derivative coupling(s).
}

\begin{document}

\maketitle

\section{Introduction}

Despite a huge success of Einstein's GR starting from the observation of Mercury perihelion to black hole observation along with the detection of gravitational waves, it has various limitations and there have been many modifications beyond Einstein's gravity to address satisfactory resolution to these  limitations. From theoretical perspective,  Einstein's GR begins with the assumption that the connection (associated with the spacetime metric) is  symmetric in its lower indices, which in turn leads to a torsion free spacetime.
Although so far  there is no perceptive experimental signature in favour of space-time torsion , however  there is no compelling reason either to assume the absence of non-symmetric part in the affine connection.
Moreover  the Einstein-Hilbert action contains only the $linear~power$ of Ricci scalar, while the diffeomorphism symmetry admits various combinations of spacetime curvature in the gravitational action. Thus one may argue that the Einstein-Hilbert action does not include all allowed terms obeying  the full diffeomorphism symmetry of the underlying theory. It is well known that Einstein's GR face various challenges in explaining various evolutionary phase of the universe, encounters the  singularity problem (also known as the Big-Bang singularity) and faces unavoidable divergences in the ultra-violet regime leading to the so called quantum gravity problem.

In order to relax the formal assumption of symmetric connection, the first attempt was made by Cartan to include the torsion in spacetime background, which with further additions was named Einstein-Cartan theory \cite{Hehl:1976kj,Shapiro:2001rz}. In particular, it was showed that the spin angular momentum of the matter field(s) can act as a source of spacetime torsion, just like the energy is the source of spacetime curvature. In this regard, there had been a plethora of work  to investigate various gravitational aspects of massless Kalb-Ramond field which is a spin one  two rank antisymmetric tensor field \cite{duff,kalb,ssg_KR1,ssg_KR2,ssg_KR3,biswarup_KR,sumanta_KR,Aashish:2019ykb,Aashish:2019zsy,Elizalde:2018rmz,Elizalde:2018now,Paul:2020duu,Paul:2022mup}. Regarding the other limitation, it is well known that Einstein–Hilbert action can be generalized by adding higher order curvature terms which naturally arise from diffeomorphism property of the action. Such terms also have their origin in String theory due to quantum corrections. $f(R)$ \cite{Nojiri:2010wj,Nojiri:2017ncd,Capozziello:2011et}, Gauss-Bonnet (GB) \cite{Nojiri:2005jg,Nojiri:2005am} or more generally Lanczos-Lovelock gravity are some of the candidates in higher curvature gravitational theory. In general inclusion of higher curvature terms in the action leads to the appearance of ghost from higher derivative terms resulting into Ostragradsky instability. The Gauss–Bonnet model (a special case of Lanczos–Lovelock model) is however free of this instability due to appropriate choice of various quadratic combinations of Riemann tensor, Ricci tensor and curvature scalar. In contrast to GB model $f(R)$ gravity model however contains higher curvature terms consisting only of the scalar curvature R. Once again just as GB model, certain classes of $f(R)$ gravity models are free from ghost-like instability. In general $f(R)$ model can be mapped into a scalar–tensor theory at the action level by a conformal transformation of the metric \cite{Nojiri:2010wj,Nojiri:2017ncd,Capozziello:2011et}. The issue of instability of the original $f(R)$ model is now reflected in the form of the kinetic and potential terms of the scalar field in the dual scalar–tensor model, where the potential will have a stable minimum and a kinetic term with proper signature. The $f(R)$ gravity theory earned the most attention in the arena of higher curvature gravitational theories, as the $f(R)$ theory can naturally unify the early inflationary phase of the universe with the dark energy era \cite{Nojiri:2010wj,Nojiri:2017ncd,Capozziello:2011et,Artymowski:2014gea,
Nojiri:2003ft,Odintsov:2019mlf,Johnson:2019vwi,Pinto:2018rfg,Odintsov:2019evb,Nojiri:2019riz,Nojiri:2019fft,Lobo:2008sg,
Gorbunov:2010bn,Li:2007xn,Odintsov:2020nwm,Odintsov:2020iui,Appleby:2007vb,Elizalde:2010ts,Cognola:2007zu}. Various cosmological and astrophysical aspects of higher curvature theories are explored in \cite{Nojiri:2010wj,Nojiri:2017ncd,Capozziello:2011et,Artymowski:2014gea,
Nojiri:2003ft,Odintsov:2019mlf,Johnson:2019vwi,Pinto:2018rfg,Odintsov:2019evb,Nojiri:2019riz,Nojiri:2019fft,Lobo:2008sg,
Gorbunov:2010bn,Li:2007xn,Odintsov:2020nwm,Odintsov:2020iui,Appleby:2007vb,Elizalde:2010ts,Cognola:2007zu,Li:2007jm,Carter:2005fu,Nojiri:2019dwl,Odintsov:2022unp,Bamba:2020qdj,Makarenko:2016jsy,delaCruzDombriz:2011wn,Chakraborty:2018scm,
Kanti:2015pda,Kanti:2015dra,Saridakis:2017rdo,Cognola:2006eg}.

As mentioned above that without spacetime torsion, $f(R)$ gravity action can be mapped to a minimally coupled scalar-tensor theory in the Einstein frame, where the scalar field appears with a potential that depends on the form of $f(R)$ under consideration. However one may expect that the scenario changes in presence of spacetime torsion, i.e., if the affine connection is considered to have an antisymmetric part. It is therefore crucial to explore whether the torsion included modified $f(R)$ model exhibits a similar dual structure 
in the form of scalar-tensor theory and how the space-time torsion couples itself with scalar degrees of freedom which originates from the higher derivative part of the action.
This is the motivation of our present work, in particular, we want to address the correspondence between the Jordan and Einstein frame $f(R)$ gravity in the presence of spacetime torsion. We also address the issue of Ostragadsky instability in the generalized scenario. Here we would like to mention that an equivalent scalar-tensor representation of Cartan $f(R)$ theory has been demonstrated in \cite{Inagaki:2022blm} where the authors showed that, similar to normal $f(R)$ gravity, Cartan $f(R)$ theory can also be represented to a minimally coupled scalar-tensor theory (without any antisymmetric tensor field). Actually the formalism of the current work is different than \cite{Inagaki:2022blm}, in particular, we transform the ``$f(R)$ gravity with torsion'' from Jordan to Einstein frame based on a conformal transformation of metric, unlike to \cite{Inagaki:2022blm} where the authors rewrite the Cartan $f(R)$ theory in Jordan frame itself by a different fashion without considering any conformal transformation of the metric. This makes our present scenario essentially different from the earlier ones. Throughout the current paper, a quantity with an overbar represent the same with respect to  the symmetric Christoffel connection, and a quantity with tilde refers to the Einstein frame quantity.

\section{Ricci Tensor with torsion}
As we know the torsion tensor is the antisymmetric part of the affine connection. We know the Riemann curvature tensor formula is given by
\begin{equation}
    {R^d}_{abc}=\partial_b({\Gamma^d}_{ac})-\partial_c({\Gamma^d}_{ab})+{\Gamma^e}_{ac}{\Gamma^d}_{be}-{\Gamma^e}_{ab}{\Gamma^d}_{ce}
\end{equation}
One can express the connection through metric and contorsion using metric compatibility in a unique way as
$${\Gamma^d}_{ab}={\bar{\Gamma}^d}_{ab}-{K^d}_{ab}$$
where ${\bar{\Gamma}^d}_{ab}$ is symmetric in the lower indices and is given by:
\begin{equation}
    {\bar{\Gamma}^d}_{ab}=\frac{1}{2}g^{dc}\left(\partial_ag_{eb}+\partial_bg_{ea}-\partial_eg_{ab} \right)
\end{equation}

and 

\begin{equation}\label{contorsion}
    {K^d}_{ab}=\frac{1}{2}({{T^d}_{ab}}-{{{T_a}^d}_b}-{{{T_b}^d}_a} )
\end{equation}
is called the contorsion tensor. The indices are raised and lowered by means of the metric. Here it may be mentioned that the contorsion is antisymmetric in the first two indices, i.e. $K_{abc} = -K_{bac}$, while the torsion tensor $T^{a}_{bc}$ itself is antisymmetric in the last two indices. Due to Eq.~(\ref{contorsion}), the Riemann tensor is written as,
\begin{equation*}
    \begin{split}
        {R^a}_{bcd}&={\bar{R}^a}_{bcd}-\partial_c {K^a}_{bd}+\partial_d {K^a}_{bc}-{\bar{\Gamma}^a}_{ec}{K^e}_{bd}-{\bar{\Gamma}^e}_{bd}{K^a}_{ec}\\
        &+{\bar{\Gamma}^a}_{ed}{K^e}_{bc}+{\bar{\Gamma}^e}_{bc}{K^a}_{ed}+{K^a}_{ec}{K^e}_{bd}-{K^a}_{ed}{K^e}_{bc}
    \end{split}
\end{equation*}
\begin{equation*}
\begin{split}
    R_{bd}&=\bar{R}_{bd}-\partial_a {K^a}_{bd}+\partial_d {K^a}_{ba}-{\bar{\Gamma}^a}_{ea}{K^e}_{bd}-{\bar{\Gamma}^e}_{bd}{K^a}_{ea}\\
    &+{\bar{\Gamma}^a}_{ed}{K^e}_{ba}+{\bar{\Gamma}^e}_{ba}{K^a}_{ed}+{K^a}_{ea}{K^e}_{bd}-{K^a}_{ed}{K^e}_{ba}
\end{split}
\end{equation*}
This can also be written as

\begin{equation}
     R_{bd}=\bar{R}_{bd}-\bar{\nabla}_a {K^a}_{bd}+\bar{\nabla}_d {K^a}_{ba}+{K^a}_{ea}{K^e}_{bd}-{K^a}_{ed}{K^e}_{ba}~~.
     \label{Ricci}
\end{equation}
Owing to the fact that the covariant derivatives (with overbar) can be expanded in the terms of the symmetric Christoffel connection, the above expression gives the expression for the Ricci tensor as a function of the symmetric Christoffel connections and the contorsion tensor. This turns out to be sum of the Ricci tensor built from the symmetric part of the christoffel connections and derivative of contorsion and higher order terms in it. We shall use this expression in the next section to modify the $f(R)$ action in terms of torsion and metric coefficients.

\section{$f(R)$ action with torsion in Einstein frame}
As mentioned in the introduction, we intend to map the $f(R)$ action from Jordan to Einstein frame in presence of spacetime torsion. The $f(R)$ action can be written as:
\begin{equation}
    S= \frac{1}{2\kappa^2}\int d^4x \sqrt{-g} f(R)~~.
    \label{J-action-1}
\end{equation}
Here $f(R)$ is an analytic function of Ricci scalar: $R = R_{\mu\nu}g^{\mu\nu}$ where Ricci tensor $R_{\mu\nu}$ contains the torsional part and given by Eq.(\ref{Ricci}). Moreover $g$ is the determinant of the metric $g_{\mu\nu}$ and $\kappa^2 = 8\pi G$ with $G$ being the Newton's gravitational constant. Owing to Eq.(\ref{Ricci}), the action (\ref{J-action-1}) can be shown as,
\begin{equation}
    \begin{split}
        S &= \frac{1}{2\kappa^2}\int d^4x\sqrt{-g}~f\big[g^{bd}(\bar{R}_{bd}-\bar{\nabla}_a {K^a}_{bd} \\ 
    &+\bar{\nabla}_d {K^a}_{ba}+{K^a}_{ea}{K^e}_{bd}-{K^a}_{ed}{K^e}_{ba})\big]~~.
    \label{J-action-2}
    \end{split}
\end{equation}
Moreover, by introducing an auxiliary field $A(x)$, the action (\ref{J-action-1}) can be equivalently written as,
\begin{eqnarray}
    S = \frac{1}{2\kappa^2}\int d^4x \sqrt{-g} \left[f'(A)\left(R - A\right) + f(A)\right]~~.
    \label{J-action-3}
\end{eqnarray}
The variation of this action over the auxiliary field $A(x)$ leads to $A=R$ which finally results to the original action (\ref{J-action-1}). The above action can be mapped to the Einstein frame by applying the following conformal transformation on the metric $g_{\mu\nu}$:
\begin{equation}
    g_{\mu\nu} \longrightarrow \widetilde{g}_{\mu\nu} = \Omega^2 g_{\mu\nu}~~,
    \label{conformal transformation}
\end{equation}
where $\Omega(x)$ is the conformal factor which is related to the auxiliary field as $\Omega^2 = f'(A)$, and from onwards, the quantities with an over-tilde represent that in the Einstein frame. Owing to this transformation the symmetric connection and the corresponding Ricci scalar (i.e without torsion) transform as:
\begin{equation}
    \bar{\Gamma}^\mu_{\nu\sigma}=\widetilde{\bar{\Gamma}}^\mu_{\nu\sigma}+\delta_\nu^\mu\partial_\sigma w+\delta_\sigma^\mu\partial_\nu w-\Tilde{g}_{\nu\sigma}\partial^\mu w
    \label{symmetruc-CC}
\end{equation}
and
\begin{equation}
    \bar{R}=\Omega^2(\widetilde{\bar{R}}+6\widetilde{\Box}w-6\widetilde{g}^{\mu\nu}\partial_\mu w\partial_\nu w)
    \label{symmetric-ricci}
\end{equation}
respectively, (recall that the symmetric connection and the corresponding quantities without torsion are denoted by an overbar). Here $w\equiv ln\Omega$ and $\widetilde{\Box}$ is the d'Alembertian operator formed by $\widetilde{g}_{\mu\nu}$, in particular
\begin{eqnarray}
    \widetilde{\Box}w \equiv \frac{1}{\sqrt{-\widetilde{g}}}\partial_\mu(\sqrt{-\widetilde{g}}~\widetilde{g}^{\mu\nu}\partial_\nu w)~~.
    \nonumber
\end{eqnarray}
Eq.(\ref{Ricci}), with the help of Eq.(\ref{symmetruc-CC}) and Eq.(\ref{symmetric-ricci}), leads to the transformation for Ricci scalar in presence of torsion. If $R$ and $\widetilde{R}$ symbolize the Ricci scalars in the Jordan and in the Einstein frame respectively, then they are related by,
\begin{equation}
    \begin{split}
    R=&\Omega^2(\widetilde{\bar{R}}+6\widetilde{\Box}w-6\widetilde{g}^{\mu\nu}\partial_\mu w\partial_\nu w)+ \widetilde{g}^{bd}\Omega^{2}[-\partial_a{K^a}_{bd}\\
    &-(\widetilde{\Gamma}^a_{ea}+\delta_a^e\partial_a w+\delta_a^a\partial_e w-\widetilde{g}_{ea}\partial^a w)K^e_{bd} \\
    &-(\widetilde{\Gamma}^e_{bd}+\delta_b^e\partial_d w+\delta_d^e\partial_b w-\widetilde{g}_{bd}\partial^e w){K^a}_{ea}\\ &+ \partial_d {K^a}_{ba}+(\widetilde{\Gamma}^a_{ed}+\delta_e^a\partial_d w+\delta_d^a\partial_e w-\widetilde{g}_{ed}\partial^a w){K^e}_{ba}\\
    &+(\widetilde{\Gamma}^e_{ba}+\delta_b^e\partial_a w+\delta_a^e\partial_b w-\widetilde{g}_{ba}\partial^e w){K^a}_{ed}\\
    &+{K^a}_{ea}{K^e}_{bd}-{K^a}_{ed}{K^e}_{ba}]~~.
    \end{split}
    \label{Ricci-transformation}
\end{equation}
Actually the first term of Eq.(\ref{Ricci}) transforms as per Eq.(\ref{symmetric-ricci}), and for other terms, we expand the covariant derivatives (with overbar) in terms of the symmetric Christoffel connection and then use the expression of Eq.(\ref{symmetruc-CC}). On simplifying Eq.(\ref{Ricci-transformation}), we get :
\begin{equation}
    \begin{split}
         R=&\Omega^2(\widetilde{\bar{R}}+6\widetilde{\Box}w-6\widetilde{g}^{\mu\nu}\partial_\mu w\partial_\nu w) + \Omega^{2}\big[2\partial^e w {{K^a}_{ea}}\\ 
         &+\widetilde{g}^{bd}(-\partial_a{K^a}_{bd}+\partial_d {K^a}_{ba}
         +{K^a}_{ea}{K^e}_{bd}-{K^a}_{ed}{K^e}_{ba}\\ 
         &-{\widetilde{\Gamma}^a}_{ea}{K^e}_{bd}-{\widetilde{\Gamma}^e}_{bd}{K^a}_{ea}+{\widetilde{\Gamma}^a}_{ed}{K^e}_{ba}+{\widetilde{\Gamma}^e}_{ba}{K^a}_{ed}) \big]~~.
    \end{split}
    \label{Ricci-transformation-1}
\end{equation}
In the above equation after adding and subtracting ${\widetilde{\Gamma}^e}_{ad}{K^a}_{be}$ it can be written as :
\begin{equation}
    \begin{split}
        R=&\Omega^2(\widetilde{\bar{R}}+6\widetilde{\Box}w-6\widetilde{g}^{\mu\nu}\partial_\mu w\partial_\nu w) \\
        &+ \Omega^{2}\big[2\partial^e w {{K^a}_{ea}}+\widetilde{g}^{bd}(-\bar{\nabla}_a {K^a}_{bd}\\ 
        &+\bar{\nabla}_d {K^a}_{ba}+{K^a}_{ea}{K^e}_{bd}-{K^a}_{ed}{K^e}_{ba})\big]~~.
    \end{split}
\end{equation}
Now using antisymmetry of K in the first two index and the metric compatibility property
 \footnote{The breaking of the metric compatibility means that one needs to add one more tensor to the affine connection. This term is called non-metricity. However, we will not consider the theories with non-metricity here.} the above equation boils down to:
\begin{equation}
       \begin{split}
             R&=\Omega^2\big[\widetilde{\bar{R}}+6\Tilde{\Box}w - 6\widetilde{g}^{\mu\nu}\partial_\mu w\partial_\nu w - 2\partial^e w {{K^a}_{ea}}\\ 
             &+2\bar{\nabla}_d {K^{ad}}_{a}+\Omega^{-2}{K^a}_{ea}{K^{ed}}_{d}-\Omega^{-2}{K}_{aed}{K}^{eda} \big]~~.
         \label{Ricci-transformation-2}
       \end{split}
\end{equation}
Now the 5th term in the above is a total divergent term and by using Gauss divergence theorem we can say that it vanishes at the boundary. Hence the final expression for the relation between $R$ and $\widetilde{\bar{R}}$ is:
\begin{equation}
         \begin{split}
             R&=\Omega^2\big[\widetilde{\bar{R}}+6\Tilde{\Box}w - 6\widetilde{g}^{\mu\nu}\partial_\mu w\partial_\nu w - 2\partial^e w {{K^a}_{ea}}\\ 
             &+\Omega^{-2}{K^a}_{ea}{K^{ed}}_{d}
             -\Omega^{-2}{K}_{aed}{K}^{eda} \big]~~.
         \label{Ricci-transformation-2}
         \end{split}
\end{equation}
Owing to the above expression of $\bar{R}$, along with the relation $\sqrt{-g}=\Omega^{-4}\sqrt{-\widetilde{g}}$, the action (\ref{J-action-3}) turns out to be,
\begin{eqnarray}
    S &=& \frac{1}{2\kappa^2}\int d^4x \sqrt{-\widetilde{g}}\bigg[f'(A)\Omega^{-2}\big(\widetilde{\bar{R}}+6\Tilde{\Box}w  
    -6\widetilde{g}^{\mu\nu}\partial_\mu w\partial_\nu w\nonumber\\ 
    &-& 2\partial^e w {{K^a}_{ea}}+\Omega^{-2}{K^a}_{ea}{K^{ed}}_{d}-\Omega^{-2}{K}_{aed}{K}^{eda}\big)\nonumber\\ 
    &-&\Omega^{-4}\left(Af'(A) - f(A)\right)\bigg]~~.
    \label{J-action-4}
\end{eqnarray}
The relation between $\Omega(x)$ and $A(x)$, i.e $\Omega^2 = f'(A)$, makes the coefficient of $\widetilde{\bar{R}}$ unity and thus the action in the Einstein frame is given by
\begin{eqnarray}
    S_\mathrm{E} = \frac{1}{2\kappa^2}\int d^4x \sqrt{-\widetilde{g}}\bigg[ \big( \widetilde{\bar{R}}- 6\widetilde{g}^{\mu\nu}\partial_\mu w\partial_\nu w - 2\partial^e w {{K^a}_{ea}} \nonumber\\
    + \Omega^{-2}{K^a}_{ea}{K^{ed}}_{d}-\Omega^{-2}{K}_{aed}{K}^{eda}\big) 
    - \frac{Af'(A) - f(A)}{f'(A)^2}\bigg]~~,
    \label{E-action-1}
\end{eqnarray}
where the suffix 'E' stands for the Einstein frame, and being a surface term, $\widetilde{\Box}w$ vanishes in the Einstein frame action.

Eq.~(\ref{E-action-1}) can be written in terms of all lowered indices as :
\begin{eqnarray}\label{action-N1}
    S_\mathrm{E}&=& \frac{1}{2\kappa^2}\int d^4x \sqrt{-\widetilde{g}}\bigg[ \big( \widetilde{\bar{R}}- 6\widetilde{g}^{\mu\nu}\partial_\mu w\partial_\nu w \nonumber\\ 
    &-& 2\partial_b w K_{cea} \widetilde{g}^{be} g^{ca}
    +\Omega^{-2}K_{cea}K_{bfd} g^{ca}g^{be}g^{fd}\nonumber\\
    &-&\Omega^{-2}K_{aed}K_{bfc} g^{ca}g^{be}g^{fd})\nonumber\\
    &-&
     \frac{Af'(A) - f(A)}{f'(A)^2}\bigg]~~
\end{eqnarray}

Due to a non-dynamical field, one can vary the above action with respect to $K_{abc}$ and get an algebraic equation in $K_{abc}$.
\begin{eqnarray}
    \delta S_\mathrm{E} &=& \frac{1}{2\kappa^2}\int d^4x \sqrt{-\widetilde{g}}\bigg[ -2\partial_b w \delta K_{cea} \widetilde{g}^{be} g^{ca}\nonumber\\&+&\Omega^{-2}(\delta K_{cea}K_{bfd}+K_{cea}\delta K_{bfd})g^{ca}g^{be}g^{fd}\nonumber\\
    &-&\Omega^{-2}(\delta K_{aed}K_{bfc}+K_{aed}\delta K_{bfc}) g^{ca}g^{be}g^{fd})\big) \bigg]~~,
\end{eqnarray}
\begin{eqnarray}
    \delta S_\mathrm{E} &=& \frac{1}{2\kappa^2}\int d^4x \sqrt{-\widetilde{g}} ~~\delta K_{jkl}\bigg[ -2\partial^e w {\delta^k}_e g^{jl}\nonumber\\
    &+&\Omega^{-2}K_{bfd}g^{jl}g^{bk}g^{fd}+\Omega^{-2}K_{cea} g^{ca}g^{je}g^{kl}\nonumber\\
    &-&\Omega^{-2}K_{bfc} g^{cj}g^{bk}g^{fl}-\Omega^{-2}K_{aed} g^{al}g^{je}g^{kd}\big) \bigg]~~,
\end{eqnarray}
hence the equation of motion becomes:
\begin{equation}
    2\partial^k w g^{jl}
    =-\Omega^{-2}{K^{kd}}_{d}g^{jl} -\Omega^{-2}{K^{aj}}_{a}g^{kl}+K^{klj}+K^{ljk}
\end{equation}
Multiplying both sides with $g_{jl}$ we get:

\begin{equation}\label{sol1}
    {K^a}_{ea}=4 \partial_e w~.
\end{equation}
Eq.~(\ref{sol1}) immediately leads to the solution of $K_{abc}$ as follows,
\begin{eqnarray}\label{sol2}
K_{abc} = \left(\frac{4}{3\Omega^2}\right)\left\{\widetilde{g}_{ac} \partial_{b}w - \widetilde{g}_{bc} \partial_{a}w\right\} + \partial_{[b}B_{ac]}
\end{eqnarray}
where $B_{[ac]}$ is a two rank antisymmetric tensor field, such that $B_{[ac]}g^{ac} = 0$, and note that the right hand side of Eq.~(\ref{sol2}) respects the antisymmetric nature of $K_{abc} = -K_{bac}$. Due to such antisymmetric property of $B_{[ac]}$, the solution in Eq.~(\ref{sol1}) can be easily obtained from the other one. Eq.~(\ref{sol2}) argues that the d.o.f of the contorsion tensor is encapsulated within a scalar field ($w$) and a 2-rank antisymmetric tensor field ($B_{[ac]}$). Using the above solution of $K_{abc}$ from Eq.~(\ref{sol2}) to Eq.~(\ref{action-N1}) along with a little bit of simplification yields the final form of the action in the Einstein frame as :

\begin{eqnarray}
    S_\mathrm{E} &=& \frac{1}{2\kappa^2}\int d^4x \sqrt{-\widetilde{g}}\bigg[ \widetilde{\bar{R}}- 2\widetilde{g}^{ef}\partial_e w\partial_f w\nonumber\\
&-&\Omega^{4}\widetilde{g}^{ca}\widetilde{g}^{be}\widetilde{g}^{fd}\partial_{[e} B_{ad]}\partial_{[f} B_{bc]}\nonumber\\ &-& \frac{Af'(A) - f(A)}{f'(A)^2}\bigg]~~,
    \label{E-action-1}
\end{eqnarray}
where we use $B_{[ac]}g^{ac} = 0$. It may be noted that neither $w$ nor $B_{[ac]}$ in the action (\ref{E-action-1}) are canonical, therefore, in order to make their kinetic terms canonical, let us redefine :
\begin{equation}
    w \rightarrow \phi\equiv\frac{2ln\Omega}{\kappa}
\end{equation}
and
\begin{equation}
   B_{[ab]} \rightarrow Z_{[ab]}\equiv\frac{\Omega^2B_{[ab]}}{\kappa}
\end{equation}

In terms of canonical fields, the action in Eq.~[\ref{E-action-1}] takes the following form:

\begin{eqnarray}\label{E-action-2}
    S_\mathrm{E} &=& \int d^4x \sqrt{-\widetilde{g}}\bigg[ \frac{\widetilde{\bar{R}}}{2\kappa^2}- \frac{1}{2}\widetilde{g}^{ef}\partial_e \phi\partial_f \phi\nonumber\\ 
    &-& \frac{Af'(A) - f(A)}{f'(A)^2} - \frac{1}{2}\partial_{[e}Z_{ad]}\partial^{[e}Z^{ad]}\nonumber\\ 
    &-&\frac{\kappa}{2}\widetilde{g}^{ac}\widetilde{g}^{be}\widetilde{g}^{fd}\left(\partial_{[e}\phi~Z_{ad]}\partial_{[f}Z_{bc]} + \partial_{[f}\phi~Z_{bc]}\partial_{[e}Z_{ad]}\right)\nonumber\\ 
    &+& \frac{\kappa^2}{2}\widetilde{g}^{ac}\widetilde{g}^{be}\widetilde{g}^{fd}\partial_{[e}\phi~Z_{ad]}\partial_{[f}\phi~Z_{bc]}
    \bigg]~,
\end{eqnarray}
where the lower and upper indices are with respect to $\widetilde{g}_{ab}$, and $\widetilde{\bar{R}}$ is the Ricci scalar in Einstein frame formed by the symmetric affine connection. The above action resembles with a scalar-tensor action along with a 2-rank massless antisymmetric tensor field ($Z_{[ab]}$), where the scalar field gets non-minimally coupled with the 2-rank tensor field. Note that $\phi(x)$ acts as a scalar field with the potential $V(A(\phi)) = \frac{Af'(A) - f(A)}{f'(A)^2}$, and the antisymmetric tensor field carries the signature of the spacetime torsion. Therefore in the context of $f(R)$ gravity in presence of spacetime torsion, the higher curvature d.o.f manifests itself as a scalar field d.o.f which couples with the massless antisymmetric tensor field (or equivalently, with the torsion field) through derivative kind of coupling. It is important to note that the scalar field and the torsion do not propagate independently, actually the torsion field gets a source term from $\phi(x)$. Clearly such coupling between the scalar and the torsion fields introduces new three or four point vertices which may have interesting phenomenological implications both in cosmology as well as in particle physics. Here it is important to note that the 3-point interaction vertex between the $\phi(x)$ and the $Z_{[ab]}$ contain the factor $\kappa$, while the 4-point vertex gets suppressed by $\kappa^2$ (this can also be understood from dimensional analysis). Therefore the interaction that can give the most significant effects is given by the 3-point interaction between $\phi$ and $Z_{[ab]}$. Being a derivative coupling, it is supposed to have significant effects during the early universe cosmological phenomena where the energy scale of the universe is of $\sim 10^{-3}$ order of the Planck scale. In order to demonstrate the cosmological implications of the interaction terms, let us go through the Friedmann equations for the Einstein action (\ref{E-action-2}). For this purpose, we consider a spatially flat FLRW metric in the Einstein frame, as follows:
\begin{eqnarray}
 d\Tilde{s}^2 = -dt^2 + a^2(t)d\textbf{x}^2~~,
 \label{rev-1}
\end{eqnarray}
where $t$ is the cosmic time and $a(t)$ represents the scale factor of the universe. With the above metric along with the consideration that $\phi$ and $Z_{ab}$ depend only on the cosmic time, the corresponding Friedmann equations turn out to be (see the Appendix for the detailed derivation),
\begin{eqnarray}
    \frac{3}{2\kappa^2}\bigg(\frac{\Dot{a}}{a}\bigg)^2&=&\frac{1}{4}(\partial_t\phi)^2+\frac{V(\phi)}{2}\nonumber\\
    &-&\frac{1}{3}\partial_tZ_{ad}\partial^tZ^{ad}
    -\frac{5\kappa}{6}\partial_t\phi Z_{ad}\partial^tZ^{ad}\nonumber\\
    &+&\frac{5\kappa^2}{12}\partial_t\phi Z_{ad}\partial^t\phi Z^{ad}
    \label{rev-2}
\end{eqnarray}
and
\begin{eqnarray}
    \frac{3}{\kappa^2}\left[\frac{\Ddot{a}}{a}+\bigg(\frac{\Dot{a}}{a}\bigg)^2\right]&=& 2V(\phi)-\frac{1}{2}(\partial_t\phi)^2\nonumber\\&-&\frac{1}{2}\partial_tZ_{ij}\partial^tZ^{ij}+\kappa\partial_t\phi Z_{ij}\partial^tZ^{ij}\nonumber\\&-&\frac{\kappa^2}{2}\partial_t\phi Z_{ij}\partial^t\phi Z^{ij}
    \label{rev-3}
\end{eqnarray}
respectively, where an overdot symbolizes $\frac{d}{dt}$. Clearly the usual cosmological field equations for a scalar-tensor theory can be recovered from the above two in absence of the antisymmetric tensor field. It is evident that the antisymmetric tensor field couples with the scalar field through its velocity term, in particular, with $\dot{\phi}$. Such a coupling should have significant effects during the early stage of the universe (particularly before the BBN), here we list some of the possibilities:
\begin{itemize}
 \item In a typical inflationary context, at the beginning when the scalar field follows the slow roll condition(s), $\dot{\phi}$ remains negligible and hence the interaction of the antisymmetric field with the scalar field is not that significant. In that case, the Hubble parameter is dominated solely by the potential term $V(\phi)$ that leads to a de-Sitter inflation. However near the end of inflation the scalar field gets its acceleration, and as a result, the coupling between the antisymmetric and the scalar fields becomes considerable through $\dot{\phi}$, which in turn affects the end stage of inflation. In particular, Eq.~(\ref{rev-2}) and Eq.~(\ref{rev-3}) indicate that the antisymmetric field produces a deceleration effect on the cosmic expansion of the universe (this can be realized as Eq.~(\ref{rev-2}) and Eq.~(\ref{rev-3}) lead to $\Ddot{a} \propto -(\Dot{Z}_{ab})^2$ in absence of the scalar field), and thus the presence of $Z_{ab}$ along with its couplings with $\phi$ makes the inflation longer compared to the case where the antisymmetric field is absent.

 \item After the end of inflation, the universe enters to a reheating stage when the scalar field generally decays to other forms of matter fields through its couplings. Therefore due to presence of the interaction terms between $\phi$ and $Z_{ab}$, the reheating scenario in the present context also results to the production of $Z_{ab}$ along with the radiation fluid from the scalar field energy density. Thus the universe, from the end of the reheating to the BBN $\sim 10^{-2}\mathrm{GeV}$, passes through an era dominated by the antisymmetric tensor field and the radiation fluid. Such a cosmological era should have interesting effects on the propagation of primordial GWs and even to the dark matter scenario where the antisymmetric field may act as the dark matter candidate.

 \item The massless rank-2 antisymmetric tensor field may be identified with the Kalb-Ramond (KR) field. The minimally coupled scalar-tensor theory along with a KR field (i.e., where there is no coupling between the scalar field and the KR field) proves to be useful for explaining various cosmic phenomena during early universe, as showed by some of our authors in \cite{Elizalde:2018rmz,Paul:2022mup}. In particular, the Kalb-Ramond is able to trigger a bouncing universe or an inflationary universe depending on the initial conditions (see \cite{Elizalde:2018rmz,Paul:2022mup}). The intriguing effects of the minimally coupled scalar-tensor-KR theory, as shown in \cite{Elizalde:2018rmz,Elizalde:2018now,Paul:2020duu,Paul:2022mup}, points the larger importance of the action (\ref{E-action-2}) where the scalar field gets coupled with the torsion field.
\end{itemize}
These indicate the immense interest of the action in Eq.~(\ref{E-action-2}) in the cosmological sector starting from inflation to reheating stage. However the detailed study of the possibilities mentioned above are outside the focus of the present paper, and are expected to study in some near future works.

Thus as a whole, ``$f(R)$ gravity with a non-zero antisymmetric part of affine connection'' in the Jordan frame is dual to an action in Einstein frame represented by a ``non-minimally coupled scalar-tensor + a massless rank-2  antisymmetric tensor field theory with totally symmetric affine connection''. It is well known that a normal $f(R)$ theory, in absence of any antisymmetric part of the affine connection, is dual to a minimally coupled scalar-tensor theory in the Einstein frame where the scalar field potential is fixed by the form of $f(R)$ under consideration. However in the present work, we find that the scenario gets different when the affine connection in the Jordan frame has an antisymmetric part which may be identified with the spacetime torsion, i.e. in the context of an Einstein-Cartan $f(R)$ theory. It turns out that $f(R)$ gravity with spacetime torsion in Jordan frame can be recast to a non-minimally coupled scalar-tensor theory with a rank-2 massless antisymmetric tensor field and with symmetric affine connection in the Einstein frame, in particular, by the action (\ref{E-action-2}) where the antisymmetric tensor field couples with the scalar field through derivative coupling. Therefore we may argue that the extra degrees of freedom arising from the antisymmetric part of the affine connection in the Jordan frame shows as a massless rank-2 antisymmetric tensor field in the Einstein frame. This is a new result that generalizes the duality between Jordan and Einstein frames in presence of spacetime torsion.

\section{Summary}
In summary, we address the correspondence between the Jordan and the Einstein frame $f(R)$ gravity in presence of spacetime torsion. Without any torsion, it is well known that $f(R)$ gravity action can be mapped to a $minimally$ coupled scalar-tensor theory in the Einstein frame, where the scalar field appears with a potential that depends on the form of $f(R)$ under consideration. However the scenario changes with spacetime torsion, in particular, if the affine connection is considered to have an antisymmetric part. It turns out that after the inclusion of torsion, the $f(R)$ action can be recast to a $non-minimally$ coupled scalar-tensor theory with a 2-rank massless antisymmetric tensor field in the Einstein frame, where the scalar field gets coupled with the antisymmetric field. Actually the antisymmetric field carries the signature of the spacetime torsion. Therefore, interestingly, the scalar field (coming from the higher curvature d.o.f) and the torsion field do not propagate independently, actually the torsion field gets a source term from the scalar field. Furthermore, such interaction between the scalar and the torsion fields comes with a derivative coupling, and thus introduces a momentum dependent interaction vertex factor --- which may have interesting phenomenological implications both in cosmology as well as in particle physics. In this regard, it may be mentioned that the 3-point interaction vertex between the $\phi(x)$ and the $Z_{[ab]}$ contain the factor $\kappa$, while the 4-point vertex gets suppressed by $\kappa^2$ (this can also be understood from dimensional analysis). Therefore the interaction that can give the most significant effects is given by the 3-point interaction between $\phi$ and $Z_{[ab]}$.

Thus as a whole, the Einstein frame action of $f(R)$ gravity with spacetime torsion is given by a non-minimally coupled scalar-tensor theory with a rank-2 massless antisymmetric tensor field and with a totally symmetric affine connection, in particular, by the action (\ref{E-action-2}) where the antisymmetric tensor field and the scalar field couple through derivative coupling. Such interactions should have some interesting cosmological implications (some of the possibilities are discussed after Eq.~(\ref{rev-3})) that are expected to study in some future work.

\section{Appendix}\label{sec-app}

As the variation of the action (\ref{E-action-2}) with respect to the metric $\Tilde{g}^{\mu\nu}$ (let us do it term by term):\\
Variation of the first term of the action (\ref{E-action-2}):
\begin{eqnarray*}
     \delta(\sqrt{-\widetilde{g}}\widetilde{R}_{\mu\nu}\widetilde{g}^{\mu\nu})=\left(\widetilde{R}_{\mu\nu}-\frac{1}{2}\widetilde{g}_{\mu\nu}\widetilde{R}\right)\delta\widetilde{g}^{\mu\nu}\sqrt{-\widetilde{g}}~~.
\end{eqnarray*}
Variation of the second and third terms of the action (\ref{E-action-2}):
\begin{eqnarray*}
    \delta\Big(\sqrt{-\widetilde{g}}\frac{1}{2}\widetilde{g}^{\mu\nu}\partial_{\mu}\phi \partial_{\nu}\phi - V(\phi) \Big)=\sqrt{-\widetilde{g}}\Big[\frac{1}{2}\partial_{\mu}\phi \partial_{\nu}\phi\\-\frac{1}{4}\partial_{\omega}\phi \partial^{\omega}\phi \widetilde{g}_{\mu\nu} -  V(\phi)\widetilde{g}_{\mu\nu}\Big]\delta\widetilde{g}^{\mu\nu}~~.
\end{eqnarray*}
Variation of the fourth term of the action (\ref{E-action-2}):
\begin{eqnarray*}
    \delta\left( \frac{1}{2}\partial_{[e}Z_{ad]}\partial^{[e}Z^{ad]}\widetilde{g}^{ac}\widetilde{g}^{be}\widetilde{g}^{fd}\sqrt{-\widetilde{g}}\right)\\=\frac{3}{2}\sqrt{-\widetilde{g}}\partial_{[e]}Z_{a\mu]}\partial^{[e]}Z^{a\sigma]}\widetilde{g}_{\sigma\nu}\delta\widetilde{g}^{\mu\nu}\nonumber\\
    -\frac{1}{2}\delta\widetilde{g}^{\mu\nu}\sqrt{-\widetilde{g}}\partial_{[e}Z_{ad]}\partial^{[e}Z^{ad]}~~.
\end{eqnarray*}
Variation of the fifth (or sixth) term of the action (\ref{E-action-2}):
\begin{eqnarray*}
    \delta\left(-\kappa\sqrt{-\widetilde{g}}\widetilde{g}^{ac}\widetilde{g}^{be}\widetilde{g}^{fd}\partial_{[e}\phi Z_{ad]}\partial^{[e}Z^{ad]}\right)\\=\frac{\kappa }{2}\sqrt{-\widetilde{g}}\widetilde{g}_{\mu\nu}\delta\widetilde{g}^{\mu\nu}\partial_{[e}\phi Z_{ad]}\partial^{[e}Z^{ad]} \nonumber\\
    - 3\kappa\sqrt{-\widetilde{g}}\widetilde{g}_{\sigma\nu}\delta\widetilde{g}^{\mu\nu}\partial_{[e}\phi Z_{a\mu]}\partial^{[e}Z^{a\sigma]}~~.
\end{eqnarray*}
Variation of the seventh term of the action (\ref{E-action-2}):
\begin{eqnarray*}
    \delta\left(\frac{\kappa^2}{2}\sqrt{-\widetilde{g}}\widetilde{g}^{ac}\widetilde{g}^{be}\widetilde{g}^{fd}\partial_{[e}\phi~Z_{ad]}\partial_{[f}\phi~Z_{bc]}\right)\\
    = 3\frac{\kappa^2}{2} \sqrt{-\widetilde{g}}\partial_{[e}\phi~Z_{a\mu]}\partial^{[e}\phi~Z^{a\sigma]}\widetilde{g}_{\sigma\nu}\delta\widetilde{g}^{\mu\nu} \nonumber\\
    -\frac{\kappa^2}{4}\sqrt{-\widetilde{g}}\partial_{[e}\phi~Z_{ad]}\partial^{[e}\phi~Z^{ad]}\widetilde{g}_{\mu\nu}\delta\widetilde{g}^{\mu\nu}~~.
\end{eqnarray*}
Using all these terms now we can apply the least action principle to obtain the gravitational equation of motion corresponding to the Einstein action (\ref{E-action-2}), and is given by:

\begin{equation}
    \begin{split}
        \frac{1}{2\kappa^2}\widetilde{G}_{\mu\nu}&=\frac{1}{2}\partial_\mu\phi\partial_\nu\phi-\frac{1}{4}\partial_{\omega}\phi \partial^{\omega}\phi \widetilde{g}_{\mu\nu}-\frac{3}{2}\partial_{[e}Z_{a\mu]}\partial^{[e}Z^{a\sigma]}\widetilde{g}_{\sigma\nu} \\
    &+\frac{1}{2}\widetilde{g}_{\mu\nu}\partial_{[e}Z_{ad]}\partial^{[e}Z^{ad]}-\frac{\kappa }{2}\widetilde{g}_{\mu\nu}\partial_{[e}\phi Z_{ad]}\partial^{[e}Z^{ad]}  \\
    &+ 3\kappa\widetilde{g}_{\sigma\nu}\partial_{[e}\phi Z_{a\mu]}\partial^{[e}Z^{a\sigma]}+\frac{\kappa^2}{4}\partial_{[e}\phi~Z_{ad]}\partial^{[e}\phi~Z^{ad]}\widetilde{g}_{\mu\nu} \\
    &-3\frac{\kappa^2}{2} \partial_{[e}\phi~Z_{a\mu]}\partial^{[e}\phi~Z^{a\sigma]}\widetilde{g}_{\sigma\nu}+\frac{V(\phi)\widetilde{g}_{\mu\nu}}{2}~~.
    \label{rev-4}
    \end{split}
\end{equation}

Eq.~(\ref{rev-4}) results to the Friedmann Eq.~(\ref{rev-2}) and Eq.~(\ref{rev-3}) with the consideration of a spatially flat FLRW metric.

\end{document}